\documentclass[aps,pre,twocolumn,longbibliography]{revtex4-2}

\usepackage[T1]{fontenc}
\usepackage[utf8]{inputenc}
\usepackage{lmodern}
\usepackage{graphicx}
\usepackage{amsmath,amssymb,mathtools}

\usepackage[caption=false]{subfig}

\usepackage{array}
\usepackage[table]{xcolor}
\usepackage{tabularx}

\usepackage{algorithm}
\usepackage[noend]{algpseudocode}

\usepackage[colorlinks=true,linkcolor=blue,citecolor=blue,urlcolor=blue]{hyperref}

\graphicspath{{figures/}}

\usepackage[caption=false]{subfig}

\usepackage{footmisc}

\usepackage[colorlinks=true,linkcolor=blue,citecolor=blue,urlcolor=blue]{hyperref}

\graphicspath{{figures/}}

\newcommand{\beq}{\begin{equation}}
\newcommand{\eeq}{\end{equation}}
\newcommand{\be}{\begin{equation}}
\newcommand{\ee}{\end{equation}}

\newcommand{\beqn}{\begin{eqnarray}}
\newcommand{\eeqn}{\end{eqnarray}}

\begin{document}

\title{Equilibrium and dynamics of a three-state opinion model on a network of networks}

\author{Irene Ferri}
\affiliation{Departament de F\'isica de la Mat\`eria Condensada and Institute of Complex Systems (UBICS), Universitat de Barcelona, 08028 Barcelona, Spain}
\author{Albert D\'\i az-Guilera}
\affiliation{Departament de F\'isica de la Mat\`eria Condensada and Institute of Complex Systems (UBICS), Universitat de Barcelona, 08028 Barcelona, Spain}
\author{Hiroki Sayama}
\affiliation{Binghamton University, State University of New York, Binghamton, NY 13902, USA}
\affiliation{Waseda University, Tokyo 169-8050, Japan}

\date{\today}

\begin{abstract}

Opinion formation models typically represent each individual as a single variable. However, in practice each individual holds interconnected beliefs whose internal organization may influence collective outcomes. To explore this dependence, we study a three-state opinion model on a network of networks in which each agent has an internal belief graph and interacts with other agents through an external social graph. Each belief can take two opposite polarized states or a neutral one and a neutrality parameter tunes the relative conviction of the neutral stance. We incorporate temperature into the model to account for external social agitation and for the tolerance of internal cognitive dissonance. We explore the stationary state and dynamics of the model using analytical approaches and Monte Carlo simulations on a fully connected external social graph, with internal belief topologies given by one-dimensional chains, cliques, and star-like structures, where there is a central core belief to which all other beliefs are connected. We find that the critical temperature at which the polarized consensus destabilizes increases with the addition of more beliefs to star-like agents but saturates in the case of ring- and clique-like internal topologies. We also consider binary mixtures of agents with different internal topologies in equal proportions, showing that the interplay between agents is regime-dependent, with the dominant topology depending on the value of the neutrality parameter.

\end{abstract}

\maketitle

\section{Introduction}\label{sec:intro}
Human beings and the belief systems that support their thought and opinion formation processes are complex in themselves. Representing opinion as a single variable offers many advantages for modeling, as it is usually simple enough to be treated analytically in some limit cases and it enables the description of a wide variety of social phenomena. However, a more detailed description of each agent’s opinion can be informative and provide insight at smaller scales.

Some models in sociophysics approach this problem by treating opinion as a vector in a multidimensional space whose components need not be independent \cite{Axelrod_1997, Baumann_2021, Pham_2022, Korbel_2023}. Such representation allows to explain collective phenomena that scalar models cannot produce, such as the local convergence with global polarization observed in culture-dissemination dynamics \cite{Axelrod_1997} and the ideological phase in which polarization coexists with alignment of stances across topics \cite{Baumann_2021}.

A complementary direction in psychology represents an individual's belief system not as coordinates of an opinion vector but as a graph whose nodes are beliefs and whose edges encode their pairwise relations. This perspective traces back to mid-20th-century work in social psychology, where an individual's attitudes were already encoded as a signed graph of positive, negative, and null connections \cite{Abelson_1958}, and continues today in modern models of social cognition \cite{Dalege_2022, Vlasceanu_2023}.

Networks of networks are a particular class of multilayer graphs in which each node is itself a network, and the inner nodes of one agent’s network are connected to the inner nodes of other agents’ networks via the external nodes’ connections. This kind of graph can be represented mathematically as a graph product, offering a rich framework to model complex systems involving multiple scales \cite{Sayama_2017}. Such representations are crucial for understanding emergent behaviors in biological, technological, and social systems, where interactions at different layers play a relevant role in determining the system’s overall dynamics.

For instance, in neuroscience, the brain can be understood as a network of networks: densely connected regions that operate as functional units are, in turn, interconnected to form a complex hierarchical structure known as the connectome \cite{Sporns_2005, Davison_2022}. Similarly, in biology, protein–protein interaction networks within cells form part of larger tissue-level and organ-level networks \cite{Zitnik_2017}, and the interplay between these scales has been shown to shape the system's statistical and dynamical properties \cite{Kim_2019, Escobar_2019}.  In social networks, individuals interact and form various organizations, contributing to higher-level networks such as schools, enterprises, associations, cities, and countries \cite{Granovetter_1973}.

Recent models have treated belief formation as a coupled process between two levels: an external social network, which transmits interpersonal influence, and an internal belief structure, which favors mutually consistent configurations and avoids contradiction. External social ties expose agents to peer pressure, while internal belief relations determine which combinations of beliefs are coherent, unstable, or resistant to change. Their interaction can generate cascading belief updates, attention-dependent polarization, and collective agreement that sometimes masks persistent individual-level conflict \cite{Friedkin_2016, vanderMaas_2020, Ellinas_2017}. Statistical-physics and dynamical-systems approaches further formalize these mechanisms by distinguishing personal, social, and external sources of dissonance and by showing that opinion equilibria or oscillations can emerge through bifurcations shaped jointly by the communication graph and the belief-system graph \cite{Dalege_2025, Bizyaeva_2025}. However, they do not systematically isolate internal belief-network topology as the control variable governing collective dynamics.

The possible outcomes for the collective behavior of a system are not only determined by the distribution of connections among nodes but rather by its interplay with the dynamical process that unfolds on that complex network. \cite{Castellano_2009}. One branch of models places opinions on a continuum, including Abelson's differential formulation of attitude change and bounded-confidence dynamics in which agents interact only within a tolerance window \cite{Abelson_1964, Deffuant_2000, Deffuant_2002}. Another branch represents opinions as discrete binary spins, encompassing the voter model and Ising-like statistical physics adaptations together with the Sznajd-Weron rule of pairwise reinforcement \cite{Holley_1975, Ising_1925, Galam_1997, Sznajd_2000}.

In many real-world social systems, a substantial fraction of individuals hold neutral or centrist opinions, and their behavior can shape the collective dynamics of opinion formation. The Blume–Emery–Griffiths (BEG) model, which introduces a third state between the up and down states of binary spin models, was originally formulated to describe the helium-3 / helium-4 phase transition \cite{Blume_1971}, but has since been adapted to sociophysical settings. The model has been used for instance to characterize the statistics of opinion time series under Glauber dynamics and to recover Schelling-type segregation as a kinetically constrained limit \cite{Yang_2010, Schelling_1971, Gauvin_2010}.

In previous works \cite{Ferri_2022, Ferri_2022_2, Ferri_2023} we have studied a special case of the BEG model applied to sociological complex systems, where each agent represents an individual in a social network. The model includes a temperature parameter which can be interpreted as social agitation and can lead to a lack of opinion alignment or consensus among agents, e.g., during political conflict. Building on this background, the present work uses a network-of-networks framework where external nodes (or supernodes) represent individuals. By fixing the external social layer we focus on the role that the topology of an individual's belief system (the inner nodes) plays in achieving both social consensus and internal agreement. We systematically compare clique, ring, and star-like internal topologies with different numbers of beliefs, and use the notion of temperature both as social agitation for the external layer and as a driver of \emph{cognitive dissonance} \cite{Festinger_1957, HarmonJones_2019} for the internal one.

The paper is organized as follows. In Section~\ref{sec:model} we introduce the model, situate it in a social context, and discuss its general features. In Section~\ref{sec:finite} we determine the order parameters as a function of temperature using Monte Carlo (MC) simulations and a mean-field approximation for topologies with constant-degree nodes.
In Section~\ref{sec:zero} we analyze the zero-temperature dynamics and identify the possible absorbing states for each internal topology.
Section~\ref{sec:mix} explores mixtures of agents with different internal topologies using MC simulations, showing that the presence of agents of a different type may change the behavior observed when studied in isolation. When $\alpha = 0.85$, a value above the mean-field tricritical point $\alpha_{tc} \approx 0.80$ identified in \cite{Ferri_2022}, all systems reach neutral consensus at zero temperature. Finally, Sec.~\ref{sec:conclusions} presents our conclusions.

\begin{figure}
\centering
\includegraphics[width=\columnwidth]{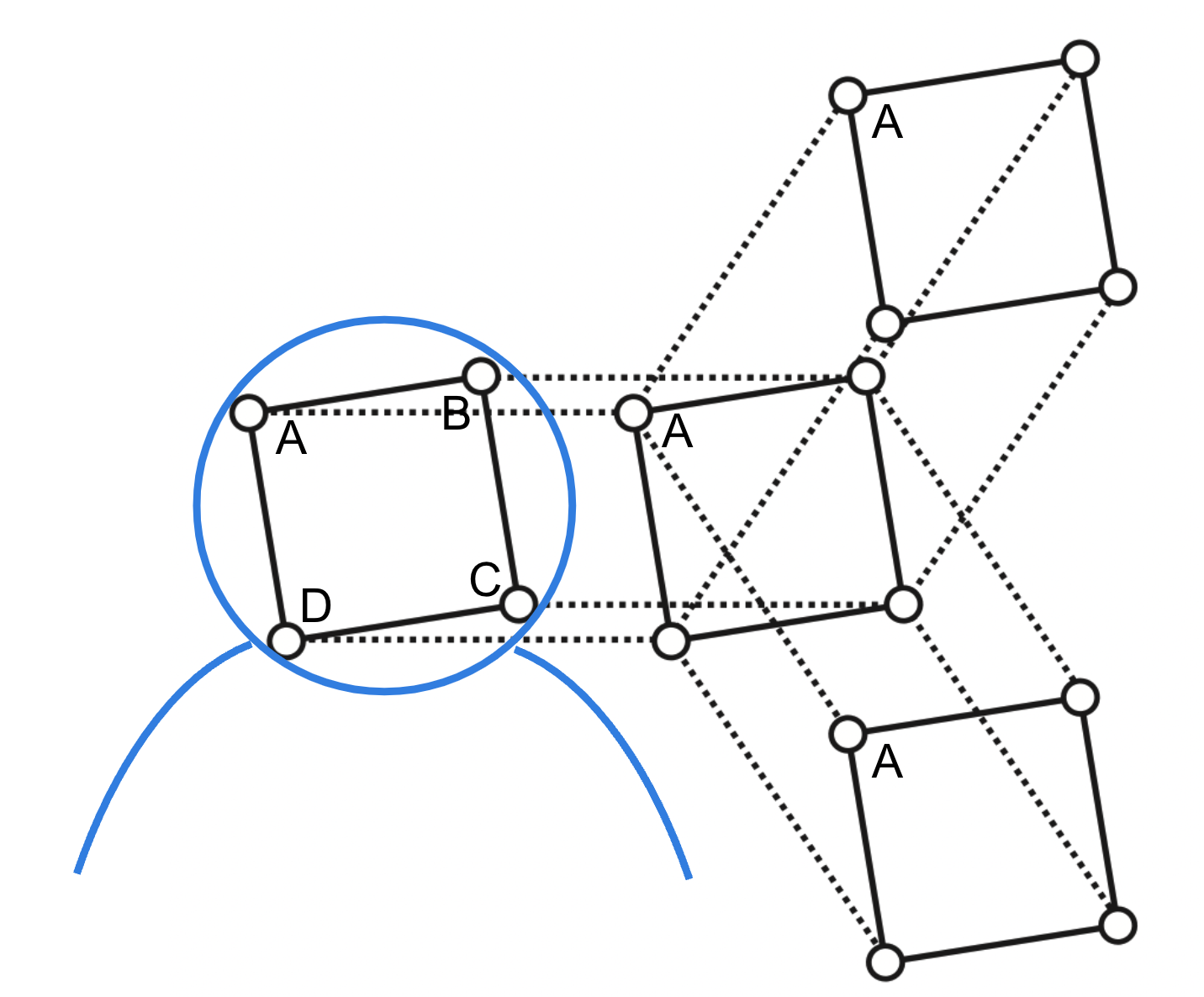}
\caption{\small{Representation of four agents with a 4-node ring-shaped internal belief topology and socially connected forming a star. Modified from \cite{Sayama_2017}.}
}
\label{Fig:cartesian_beliefs}
\end{figure}

\section{The Model}\label{sec:model}

We consider a social adjacency matrix representing the connections among individuals in a population. Each agent is a sub-system represented by an internal network of subnodes (beliefs) indexed by $\mu \in \{1,\dots,c\}$. Each belief has a topic label $\mu$ and interacts across agents only with beliefs of the same label; i.e., in a single interaction (communication act) agents discuss a specific topic. Internally, all beliefs inside an agent form a network with a given topology. The complete structure can be obtained as the Cartesian product of the external (social) and internal (belief) graphs (see Fig.~\ref{Fig:cartesian_beliefs}), which at the adjacency-matrix level corresponds to
\begin{equation}
  A \;=\; A_{\mathrm{ext}} \otimes I_c \;+\; I_N \otimes A_{\mathrm{int}},
  \label{eq:cartesian}
\end{equation}
where $A_{\mathrm{ext}}$ is the $N\times N$ social adjacency matrix, $A_{\mathrm{int}}$ is the $c\times c$ internal (belief) adjacency, $I_k$ is the $k\times k$ identity, $\otimes$ denotes the Kronecker product, and nodes are labeled by pairs $(i,\mu)$ with $i\in\{1,\dots,N\}$ and $\mu\in\{1,\dots,c\}$.

\begin{figure*}
    \centering
    \includegraphics[width=\textwidth]{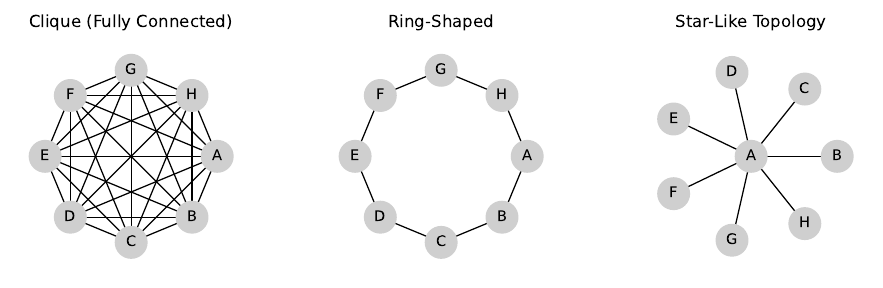}
    \caption{Illustration of the three network topologies used to model belief systems with a number of internal beliefs $c = 10$.}
    \label{fig:inner_topologies}
\end{figure*}

For the social external network, we consider a fully connected graph in order to isolate the effects of the internal belief topology from those of the social network structure. For the internal structures, we explore three distinct networks to represent different individual belief systems (see Fig. \ref{fig:inner_topologies}):
\begin{itemize}
  \item \textbf{Clique (fully connected):} proxy for a richly interconnected belief system.
  \item \textbf{Ring-shaped:} exemplifies sparse inner connections.
  \item \textbf{Star-like:} denotes individuals with a dominant core belief supported by all other belief components.
\end{itemize}

Our goal is to investigate how these internal topologies influence the consensus state of a population whose beliefs may take one of three orientations. For star-like agents, we consider the simplest case in which the central hub corresponds to the same belief topic in every agent. Allowing the identity of the hub belief to vary across agents would be a natural extension of the model, but lies outside the scope of the present study.

Inspired by magnetic spin-like models, we consider three possible states for every belief: two extreme opposite states and a neutral one. The third (neutral) state is placed symmetrically between the two polarized ones, and we quantify the transition probability toward this central state using a \emph{neutrality} parameter $\alpha$, capturing the relative degree of conviction between neutral and extremist beliefs.

Each belief state is represented by a two-dimensional vector $\bar{\mathbf{s}}_{i}^{\mu}$ that can take three orientations:
\begin{align}
\bar{\mathbf{s}}_{i}^{\mu} &= (1,\,0) \qquad &&\text{positive / in favor}, \nonumber \\
\bar{\mathbf{s}}_{i}^{\mu} &= (0,\,\alpha) \qquad &&\text{neutral / central}, \nonumber \\
\bar{\mathbf{s}}_{i}^{\mu}  &= (-1,\,0) \qquad &&\text{negative / against}, \nonumber
\end{align}
where $\alpha$ is a dimensionless parameter.

\subsection{Hamiltonian}
\begin{figure}
    \centering
    \includegraphics[width=\columnwidth]{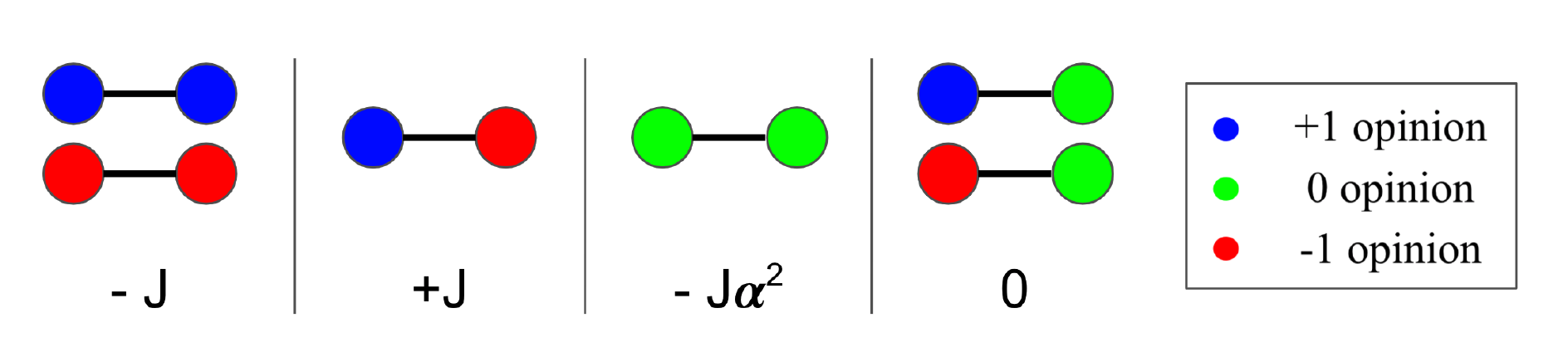}
    \caption{Contribution of a pair of interacting agents to the energy of the system.}
    \label{Fig:contributions}
\end{figure}

Motivated by homophily, the observation that individuals tend to associate with those who hold similar views \cite{McPherson_2001}, we assume that agents prefer to agree with their neighbors and, at the same time, to maintain internal coherence among their beliefs. In the absence of temperature $T$, the system tends to minimize the following cost function (Hamiltonian):
\begin{equation}
\mathcal{H}
= -\frac{J}{z_{\mathrm{ext}}}\sum_{\mu=1}^{c}\,\sum_{\langle i,j\rangle_{\mathrm{ext}}} \bar{\mathbf{s}}_{i}^{\mu}\!\cdot\!\bar{\mathbf{s}}_{j}^{\mu}
\;-\;
\frac{J}{z_{\mathrm{int}}}\sum_{i=1}^{N}\,\sum_{\langle \mu,\nu\rangle_{\mathrm{int}}} \bar{\mathbf{s}}_{i}^{\mu}\!\cdot\!\bar{\mathbf{s}}_{i}^{\nu},
\label{eq:hamiltonian}
\end{equation}
where $J>0$ is a ferromagnetic coupling (set to $J=1$ for simplicity), $z_{\mathrm{ext}}$ is the average degree of social connections (for a complete graph, $z_{\mathrm{ext}}=N-1$), and $z_{\mathrm{int}}$ is the average internal degree, which depends on the chosen belief-topology (e.g., clique: $z_{\mathrm{int}}=c-1$; ring: $z_{\mathrm{int}}=2$; star: $z_{\mathrm{int}}=2(c-1)/c$). In our notation, the first index $i$ labels agents and the second index $\mu$ labels beliefs (topics); thus, $\bar{\mathbf{s}}_{i}^{\mu}$ denotes belief $\mu$ of agent $i$. The first term couples same-topic beliefs $\mu$ across different agents (social layer), while the second couples beliefs within each agent (internal layer). Pair contributions are computed as scalar products of the two-dimensional opinion vectors (see Fig.\ref{Fig:contributions}), and the normalization constants ($z_{\mathrm{ext}}$, $z_{\mathrm{int}}$) balance the relative weight of inter-agent and intra-agent links, ensuring that both terms in $\mathcal{H}$ scale as $\mathcal{O}(Nc) $

Our goal here is not to tune the relative strength of peer pressure vs. cognitive dissonance, but to study how connection patterns in the agents’ belief networks affect macroscopic behavior. The order parameters we track are:
\begin{equation}
m = \frac{1}{Nc}\sum_{i=1}^{N}\sum_{\mu=1}^{c} \big\langle \sigma_{i}^{\mu} \big\rangle \, ,
\qquad
n_{0} = 1 - \frac{1}{Nc}\sum_{i=1}^{N}\sum_{\mu=1}^{c} \big\langle (\sigma_{i}^{\mu})^{2} \big\rangle \, ,
\label{eq:orderparams}
\end{equation}
where $\sigma_{i}^{\mu}\in\{-1,0,1\}$ denotes the first component of the opinion vector $\bar{\mathbf{s}}_{i}^{\mu}$ (negative, neutral, or positive) and $\langle \cdot \rangle$ denotes the thermal average with respect to the Boltzmann distribution. Thus $m$ measures the net polarization, i.e., the difference between the fractions of positive and negative beliefs across the system, while $n_0$ is the fraction of neutral beliefs. Order parameters can also be defined per topic $\mu\in\{A,B,C,\dots\}$ or aggregated within each agent.
For the per-agent quantities used below, we define
\begin{equation}
\begin{aligned}
E_{\mathrm{int},i} &=
-\frac{J}{z_{\mathrm{int}}}
\sum_{\langle\mu,\nu\rangle_{\mathrm{int}}}
\bar{\mathbf{s}}_{i}^{\mu}\!\cdot\!\bar{\mathbf{s}}_{i}^{\nu},\\
M_{\mathrm{int},i} &= \left|\sum_{\mu=1}^{c}\sigma_{i}^{\mu}\right|,\\
N_{0,\mathrm{int},i} &= \sum_{\mu=1}^{c}\left[1-(\sigma_{i}^{\mu})^2\right],\\
N_{+,\mathrm{int},i} &= \sum_{\mu=1}^{c}\frac{(\sigma_{i}^{\mu})^2+\sigma_{i}^{\mu}}{2}.
\end{aligned}
\label{eq:internalparams}
\end{equation}
Here $E_{\mathrm{int},i}$ is the internal contribution to the energy of agent $i$, $M_{\mathrm{int},i}$ is its absolute internal magnetization, $N_{0,\mathrm{int},i}$ counts its neutral beliefs, and $N_{+,\mathrm{int},i}$ counts its beliefs in state $+1$. In figures and discussion of distributions across agents, we omit the agent index when no ambiguity arises.

\subsection{Temperature}
Although the model favors belief alignment as the main updating mechanism, we also allow an agent $i$ to adopt a state $\bar{\mathbf{s}}_{i}^{\mu}$ that does \emph{not} minimize its local contribution to the energy, with some probability. This probability is governed by a \emph{temperature} parameter $T$, by analogy with physical systems in contact with a thermal bath. Operationally, we use conventional Metropolis dynamics, so $T$ coarse-grains all sources of randomness that may lead an agent to take a stand against their neighbors (e.g., social agitation).

For the internal network, the temperature requires a different interpretation. Here, the mechanism by which an individual may become internally misaligned is cognitive dissonance, as described by Festinger (1957) \cite{Festinger_1957}: the psychological discomfort experienced when simultaneously holding contradictory beliefs or when confronted with conflicting information. This discomfort tends to be reduced via belief revision, justification, or de-emphasizing conflicting beliefs. In this work we keep fixed the number of beliefs and internal link weights, and focus exclusively on state changes of beliefs.

We assume that both agents and beliefs are coupled to a thermal bath at temperature $T$ (in units where the Boltzmann constant $k_B\!=\!1$). In the social layer, $T$ represents a coarse-graining of exogenous fluctuations ("social agitation'') that perturb opinion formation during interpersonal communication. In the internal layer, a low temperature indicates a high resistance to cognitive dissonance (strong internal coherence), whereas a high temperature implies that individuals can sustain contradictory opinions, compromising internal coherence and reflecting a more emotionally aroused state. Since social agitation and personal emotional arousal can be related, we adopt, as a simplifying assumption, a common temperature for both layers in the baseline model. A natural extension is to assign different temperatures to the two layers (and even topic-dependent temperatures, reflecting that some topics may be more controversial than others), which implicitly reweights internal coherence versus external consensus, as explored in related work \cite{Rodriguez_2016}.

The stochastic evolution follows Metropolis Monte Carlo (MC) dynamics: at each elementary step, one belief $\bar{\mathbf{s}}_{i}^{\mu}$ is chosen uniformly at random, a new state is proposed, and the update is accepted with probability $\min\{1,e^{-\beta\Delta \mathcal{H}}\}$.
At low temperatures, depending on the underlying graphs and initial conditions, simulations may converge to the ground state of $\mathcal{H}$ or become trapped for long times in metastable configurations. Such non-equilibrium states have been observed in this model even with a single internal belief for certain external topologies, and are particularly common in modular networks; they have also been reported on Barabási–Albert and Erdős–Rényi networks with low average degree \cite{Ferri_2022, Ferri_2022_2}.

\section{Finite temperature behavior}\label{sec:finite}
In this section, we examine how the order parameters of a particular belief vary with temperature $T$ and with the number $c$ of internal beliefs for each internal topology. The order parameters defined in Eq.~\eqref{eq:orderparams} can be restricted to a single topic $\mu$ by summing only over agents, yielding the per-belief magnetization $m^{\mu}$ and neutral fraction $n_{0}^{\mu}$. We choose two values for the neutrality parameter: $\alpha = 0$, for which we expect a second-order phase transition, and $\alpha = 0.85$, which lies in the first-order transition regime for the fully connected internal graph. Our aim is to understand how the critical and tricritical points vary with system parameters and with the beliefs' topology. Since, for both the clique and the ring-like topology, all connections are symmetric for all internal nodes, we consider only belief $A$ without loss of generality. In the case of star-like individuals, we consider both $A$ (the core belief) and $B$ (one of the peripheral beliefs). Since all peripheral are topologically equivalent in the star, and each belief interacts across agents only with the same topic label, the behavior of B is representative of all peripheral beliefs.

\subsection{Analytical approximation}
Here, we extend the methodology applied in \cite{Ferri_2022} to a broader
context. We use the Weiss mean-field approximation (MFA) to build the
effective single-belief Hamiltonian for a particular belief $A$ within a
given agent. The expected value of the order parameters follows from:

\begin{equation}
\langle \bar{\mathbf{s}}^{A} \rangle
= \frac{\operatorname{tr}\!\left\{\,\bar{\mathbf{s}}^{A}\exp\!\left[\beta \bar{\mathbf{s}}^{A}\!\cdot\!(zJ\,\bar{\mathbf{m}})_{\mathrm{ext}}
+ \beta \bar{\mathbf{s}}^{A}\!\cdot\!\left(\sum_{\mu} zJ\,\bar{\mathbf{s}}^{\mu}\right)_{\mathrm{int}}\right]\right\}}
{\operatorname{tr}\!\left\{\exp\!\left[\beta \bar{\mathbf{s}}^{A}\!\cdot\!(zJ\,\bar{\mathbf{m}})_{\mathrm{ext}}
+  \beta \bar{\mathbf{s}}^{A}\!\cdot\!\left(\sum_{\mu} zJ\,\bar{\mathbf{s}}^{\mu}\right)_{\mathrm{int}}\right]\right\}},
\label{eq:MFA}
\end{equation}

where $(zJ\,\bar{\mathbf{m}})_{\mathrm{ext}} =
z_{\mathrm{ext}}\,J_{\mathrm{ext}}\,(m^{A},\,\alpha\, n_{0}^{A})$ is
the Weiss effective field from the social layer, with
$z_{\mathrm{ext}}\,J_{\mathrm{ext}} = 1$, and the sum over $\mu$ runs
over the $z_A$ internal neighbors of belief $A$.

We apply the Weiss mean-field approximation to the external (social) layer, replacing external neighbors by the average magnetization $(m^{A},\,\alpha \, n_{0}^{A})$ with effective coupling $z_{\mathrm{ext}}\,J_{\mathrm{ext}} = 1$. For the internal layer, we trace exactly over all $3^{z_A}$ configurations of the $z_A$ internal neighbors of belief $A$, treating them as independent spins coupled to $A$. The effective coupling used per neighbor in the internal trace is $J_{\mathrm{int}} = 1$
, which differs from the bare per-edge coupling $J_{\mathrm{int}} = J/z_{\mathrm{int}}$
 prescribed by the Hamiltonian. This choice compensates for the fact that the internal neighbors are traced as free, unbiased spins: in the actual system, each neighbor is biased by its own social mean field, an effect absent from the exact internal trace. Using $J_{\mathrm{int}} = J/z_{\mathrm{int}}$ with unbiased neighbors would cause internal effects to vanish as $c$ grows, incorrectly reducing the model to the single-layer case. The self-consistent solutions obtained with this approximation are in good agreement with Monte Carlo simulations (see Figs.~\ref{Fig:high_T_a0}, ~\ref{Fig:high_T_a085}).

For homogeneous-degree internal topologies, the approximation depends on the number of internal neighbors $z_A$: since $z_A = 2$ for the ring regardless of $c$, the ring predictions are independent of the number of beliefs, whereas for the clique ($z_A = c-1$) they depend on $c$ (see Table~\ref{tab:MFA}). For the star topology, where the local degree is heterogeneous, Eq.~\eqref{eq:MFA} does not distinguish between hub and peripheral beliefs. Capturing this asymmetry would require a Bethe-Peierls, or cavity/message-passing, approximation with distinct hub and leaf cavity fields. We leave that extension outside the scope of the present study and analyze star agents using MC simulations only.

\begin{table*}[t]
\centering
\caption{MFA Order Parameters for Different Cases}
\label{tab:MFA}

\begin{tabularx}{\textwidth}{| >{\centering\arraybackslash}X |}
\hline
\rowcolor{gray!30}\textbf{2 Beliefs}\\
{\small
\[
\begin{aligned}
m^{A} &= \frac{2\sinh(\beta m^{A})(2 \cosh \beta + 1)}{2\cosh(\beta m^{A})(2 \cosh \beta + 1) + \exp(\alpha^2\beta n_{0}^{A})(\exp(\alpha^2\beta) + 2)} \\
n_{0}^{A} &= \frac{\exp(\alpha^2\beta n_{0}^{A})(\exp(\alpha^2\beta) + 2)}{2\cosh(\beta m^{A})(2 \cosh \beta + 1) + \exp(\alpha^2\beta n_{0}^{A})(\exp(\alpha^2\beta) + 2)}
\end{aligned}
\]
}\\
\hline
\rowcolor{gray!30}\textbf{Triangle (Ring)}\\
{\small
\[
\begin{aligned}
m^{A} &= \frac{2\sinh(\beta m^{A}) \left [ 2 \cosh(2\beta) +4 \cosh \beta  + 3 \right] }{2\cosh(\beta m^{A})[ 2 \cosh(2\beta) +4 \cosh \beta  + 3] + \exp(\alpha^2\beta n_{0}^{A})(\exp(2\alpha^2\beta)  +  4 \exp(\alpha^2\beta) + 4)} \\
n_{0}^{A} &= \frac{\exp(\alpha^2\beta n_{0}^{A})(\exp(2\alpha^2\beta)  +  4 \exp(\alpha^2\beta) + 4)}{2\cosh(\beta m^{A})[ 2 \cosh(2\beta) +4 \cosh \beta  + 3] + \exp(\alpha^2\beta n_{0}^{A})(\exp(2\alpha^2\beta)  +  4 \exp(\alpha^2\beta) + 4)}
\end{aligned}
\]
}\\
\hline
\rowcolor{gray!30}\textbf{Clique 4 Beliefs}\\
{\small
\[
\begin{aligned}
m^{A} &= \frac{2\sinh(\beta m^{A}) \left[ 2 \cosh(3\beta) +6 \cosh (2\beta) +12\cosh \beta  +7 \right]  }{Den_4} \\
n_{0}^{A} &= \frac{\exp(\alpha^2\beta n_{0}^{A})\left[ \exp \left(3\alpha^2\beta \right) +  6\exp \left(2\alpha^2\beta \right)  +  12 \exp \left(\alpha^2\beta \right) + 8 \right]}{Den_4} \\
Den_4 &= 2\cosh(\beta m^{A}) \left[ 2 \cosh(3\beta) +6 \cosh (2\beta) +12\cosh \beta  +7 \right] + \\
&+ \exp(\alpha^2\beta n_{0}^{A})\left[ \exp \left(3\alpha^2\beta \right) + 6\exp \left(2\alpha^2\beta \right)  +  12 \exp \left(\alpha^2\beta \right) + 8 \right]
\end{aligned}
\]
}\\
\hline
\rowcolor{gray!30}\textbf{Clique 5 Beliefs}\\
{\small
\[
\begin{aligned}
m^{A} &= \frac{2\sinh(\beta m^{A}) \left[ 2 \cosh(4\beta) + 8 \cosh(3\beta) +20 \cosh (2\beta) +64\cosh \beta  +19 \right]  }{Den_5} \\
n_{0}^{A} &= \frac{2\exp(\alpha^2\beta n_{0}^{A}) \left[ \exp \left(4\alpha^2\beta \right) + 4\exp \left(3\alpha^2\beta \right) + 12\exp \left(2\alpha^2\beta \right)  +  16 \exp \left(\alpha^2\beta \right) + 8 \right]}{Den_5}  \\
Den_5 &= 2\cosh(\beta m^{A}) \left[ 2 \cosh(4\beta) + 8 \cosh(3\beta) +20 \cosh (2\beta) +64\cosh \beta  +19 \right] + \\
&+ 2\exp(\alpha^2\beta n_{0}^{A})\left[ \exp \left(4\alpha^2\beta \right) + 4\exp \left(3\alpha^2\beta \right) + 12\exp \left(2\alpha^2\beta \right)  +  16 \exp \left(\alpha^2\beta \right) + 8 \right]
\end{aligned}
\]
}\\
\hline
\end{tabularx}
\end{table*}

\subsection{Order parameters versus temperature}
Here we present results for the order parameters as a function of the temperature for different numbers of internal beliefs. We compare MFA and MC results for two values of the neutrality parameter: $\alpha=0$, which is below the tricritical point obtained in \cite{Ferri_2022}, and $\alpha=0.85$, which is above.

\smallskip
\subsubsection{Results for \texorpdfstring{{\boldmath$\alpha=0$}}{alpha=0}}

\begin{figure*}
\centering
\includegraphics[width=0.7\textwidth]{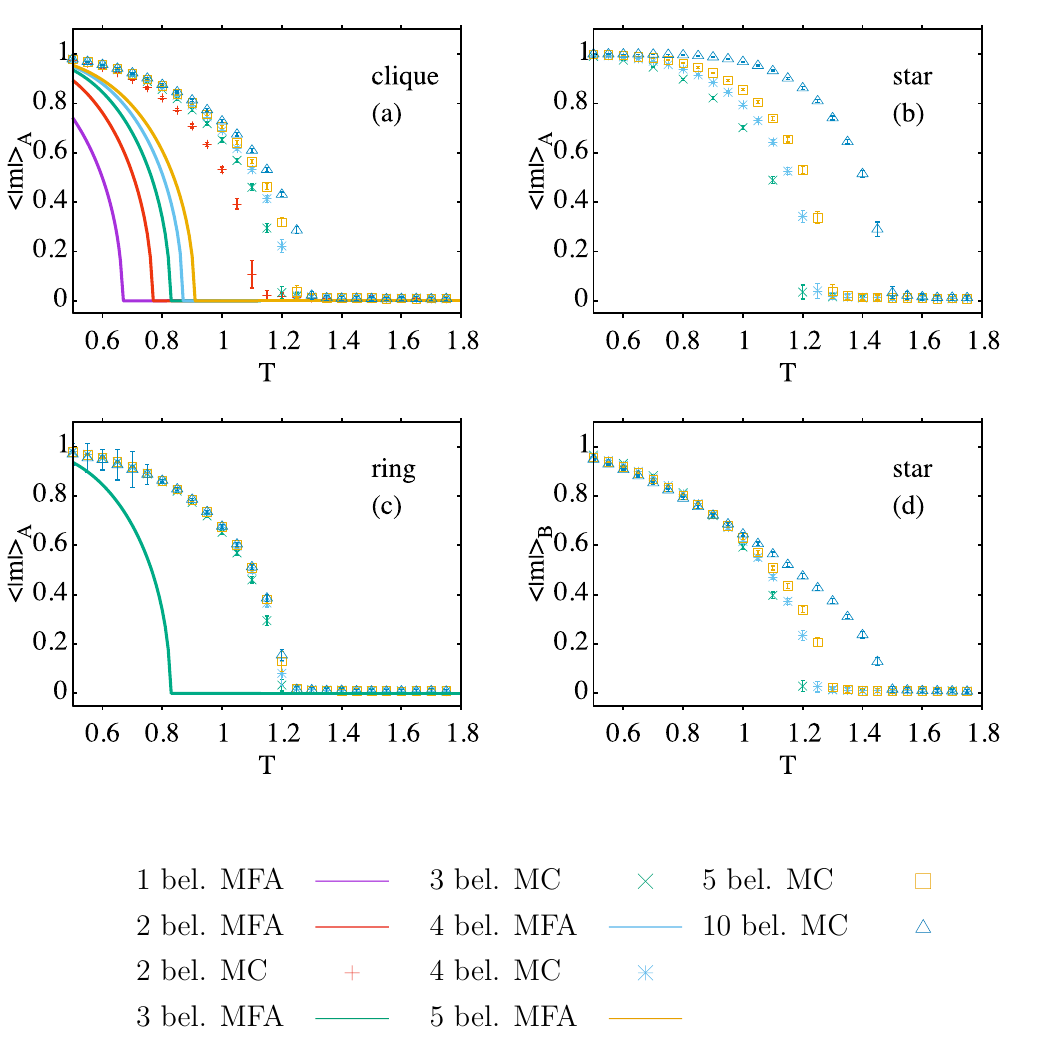}
\caption{\small{Magnetization curves for a particular belief as a function of temperature for each internal topology and different numbers of internal beliefs. Results obtained with MC simulations for a system of $N = 10^4$ agents, neutrality parameter $\alpha = 0$, averaged over $100$ repetitions with $c \cdot 10^4$ MC steps. Comparisons with MFA results are shown for the clique and the ring.}}
\label{Fig:high_T_a0}
\end{figure*}

When $\alpha=0$, all topologies exhibit a second-order phase transition for the number of internal beliefs studied. As observed for ER graphs \cite{Ferri_2022}, the MFA underestimates the critical temperature value, yet it qualitatively captures the influence of additional beliefs on it for the clique and ring internal topologies, as shown in Fig.~\ref{Fig:high_T_a0}(a),(b).

The addition of a second belief results in a considerable increase in the critical temperature $\Delta T(c=1 \to 2)$ compared to the single belief case analyzed in \cite{Ferri_2022} (see Fig.~\ref{Fig:high_T_a0}(a)). The increment caused by adding a third belief $\Delta T(c=2 \to 3)$ is smaller than $\Delta T(c=1 \to 2)$ for the triangle agents and slightly lower for the 3-node open chain agents. Further additions of beliefs do not affect the ring agents and all order parameter curves in Fig.~\ref{Fig:high_T_a0}(b) collapse, both for the MC simulations and the MFA. However, for $c=10$ the magnetization exhibits larger fluctuations at low temperature. This effect is due to the tendency of nodes in a circular topology to form clusters separated by boundaries. Although these boundaries are destined to annihilate when the number of beliefs is finite, the annihilation process significantly grows as the number of nodes in the circle increases.

On the other hand, clique and star agents continue to increase their $T_c$ when adding more beliefs to each agent, although at different rates, as shown both in Fig.~\ref{Fig:high_T_a0}(a),(c), and in Fig.~\ref{Fig:high_T_a085}.

The behavior of peripheral beliefs in star agents is different. Magnetization curves for belief $B$ regardless of the number of beliefs per agent $c$ coincide at low temperatures (see Fig.~\ref{Fig:high_T_a0}(d)), reflecting the fact that every peripheral belief has only one internal link to the central agent, regardless of the size of the internal network. However, as the critical point is approached, the correlation length grows and becomes comparable to the system size, so the behavior ceases to be purely local. As a result, the peripheral-belief curves separate and converge to the same critical temperature as belief $A$, as the system undergoes a collective transition.

Results for $n_0$ at $\alpha=0$ are not relevant, as they relate solely to thermal effects and are similar to those obtained in \cite{Ferri_2022} but with the shifted critical temperature for each case.

\begin{figure}
\centering
\includegraphics[width=0.9\columnwidth]{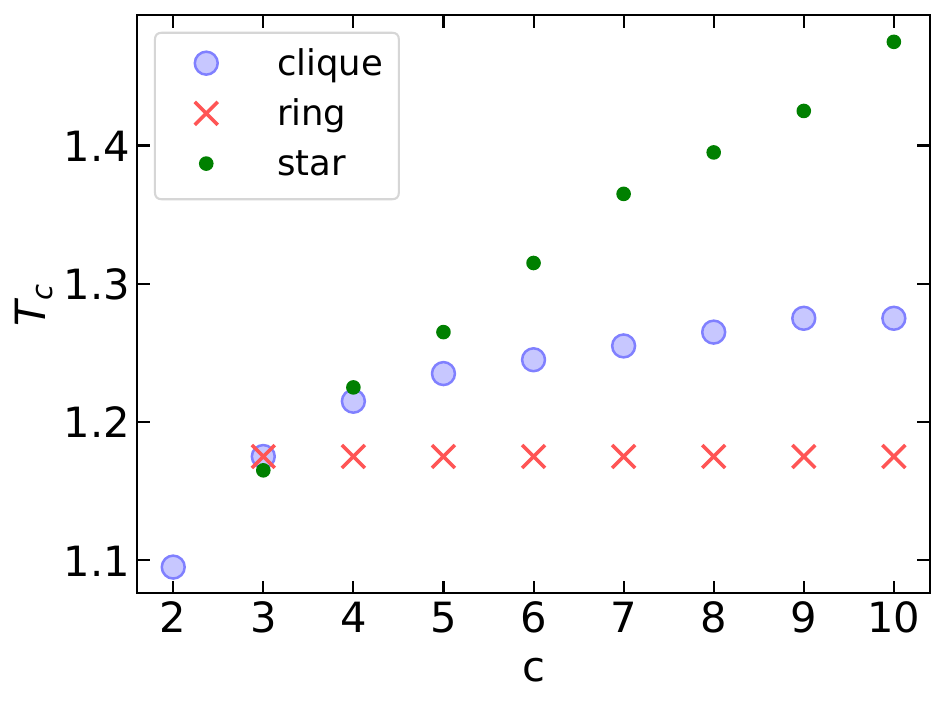}
\caption{\small{Critical temperature versus number of beliefs when $\alpha = 0$ for the clique (blue circles), the ring (red crosses), and the star-like (green dots) agent types. Results obtained averaging 100 repetitions with systems of $N = 10^4$ agents.}}
\label{Fig:phase}
\end{figure}

Fig. \ref{Fig:phase} displays the critical temperature for $\alpha=0$ as a function of the number of internal beliefs. The results were obtained using MC simulations by identifying the midpoint between the temperatures corresponding to the magnetization segment with the minimum slope.

The critical temperature for the clique increases monotonically with the number of internal beliefs within the interval $c\in[4,10]$, though the increments diminish as $c$ increases, presumably reaching a saturation point at or slightly beyond $c \approx 10$. By contrast, for star-like agents $T_c$ continues to increase over the range of $c$ explored here, approximately linearly and without the apparent saturation seen for clique agents. Peripheral beliefs act as a support to the core belief $A$, making star agents behave as ``zealots'' for $\alpha=0$ if the number of nodes in the internal network is large enough. The case with $c=3$, corresponding to a 3-beliefs open chain, deviates from this trend; in fact, $T_c(c=3)$ for the star topology is slightly lower than for the triangle, indicating that it is easier to destabilize consensus in the 3-beliefs open chain.

\smallskip
\subsubsection{Results for \texorpdfstring{{\boldmath$\alpha=0.85$}}{alpha=0.85}}

When all nodes have the same degree $k_i = z$, which holds true for both ring and clique agents, all beliefs share the same magnetization such that $m^{\mu} = m$ and the fraction of neutral beliefs $n_{0}^{\mu} = n_0$ for all $\mu \in \{A,B,C,\dots\}$. Consequently, the mean-field free-energy function $\mathcal{L}(m, n, \beta) \equiv (H_{\text{MF}} - \beta^{-1} S_{\text{MF}})/N$
simplifies to:

\begin{widetext}
\begin{equation}
\label{eq:free-energy}
\mathcal{L}(m,n,\beta)
= -\tfrac{z}{2}\!\left(m^{2}+\alpha^{2}n^{2}\right) + z\,\alpha^{2}n
+ \tfrac{1}{\beta}\!\left[
\tfrac{n+m}{2}\,\ln\!\tfrac{n+m}{2}
+ \tfrac{n-m}{2}\,\ln\!\tfrac{n-m}{2}
+ (1-n)\,\ln(1-n)
\right]
\end{equation}
\vspace{-0.6\baselineskip}
\end{widetext}

By solving the equations presented in Table~\ref{tab:MFA}, both with and without the constraint that $m=0$, and substituting the resulting solutions into Eq.~\eqref{eq:free-energy}, we determine the free energy associated with the two competing minima. The global minimum switches discontinuously between these solutions, producing a jump in the equilibrium order parameters and thus indicating a first-order transition. In the simulations, the temperature at which this occurs is close to $T=1$ for most cases, with lower values only for the triangle and for the two-belief system. The MFA underestimates the transition temperature, and this limitation is particularly evident in the two-belief case, where it predicts a lower transition temperature than in the single-belief case, contrary to the simulation results. For star agents, MC simulations show that the decay to zero magnetization occurs at a constant $T=1$ regardless of the number of beliefs and it is particularly abrupt for $c=10$.

\begin{figure*}
  \centering
  \includegraphics[width=0.7\textwidth]{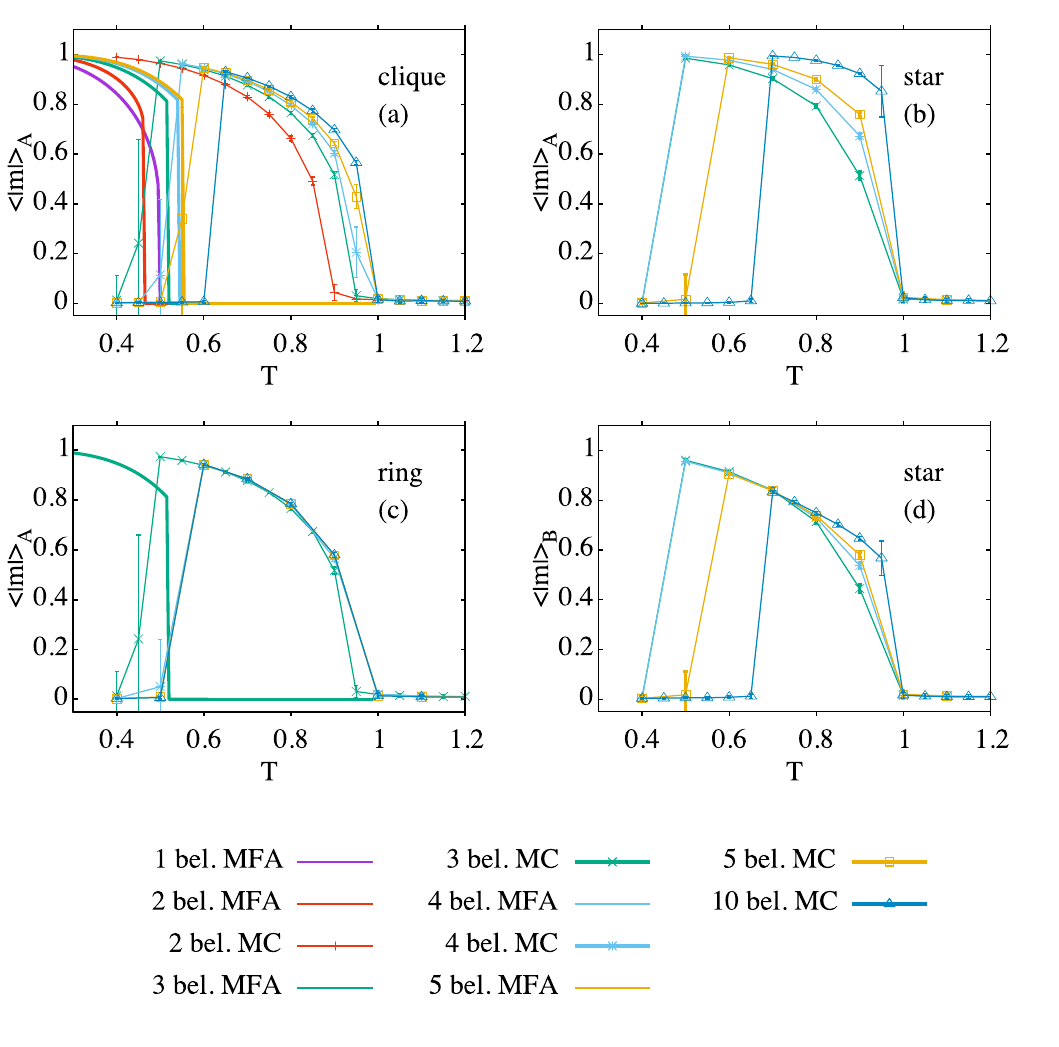}
  \caption{\small Magnetization curves for a particular belief as a function of temperature for each internal topology, and a different number of internal beliefs. Results obtained with MC simulations for a system of $N = 10^4$ agents, with a neutrality parameter $\alpha = 0.85$, and averaged over 100 repetitions. The curves are compared with MFA results for the clique and the ring topologies.}
  \label{Fig:high_T_a085}
\end{figure*}

At very low temperatures, the dynamics is unable to overcome the free-energy barrier separating the metastable neutral-consensus state from the polarized ground state, so simulations remain trapped near zero magnetization. Unlike the disordered phase at high temperature, this low-temperature near-zero magnetization is a dynamical effect. At intermediate temperatures, thermal fluctuations allow the system to cross the barrier between the neutral and polarized basins of attraction, producing the sharp increase in magnetization observed in Fig.~\ref{Fig:high_T_a085}. When the temperature is increased further, the system reaches the first-order transition described above: the polarized state loses stability and the magnetization drops again toward the disordered phase. Thus, the nonmonotonic shape of the MC curves reflects two distinct mechanisms, a dynamical escape from the neutral basin at lower temperature and an equilibrium loss of collective order at higher temperature.

The location of the dynamical jump depends on the internal topology and on the number of internal beliefs. For clique agents, this dynamical transition becomes sharper and occurs at higher temperatures as $c$ increases. The transition-temperature rise from $4$ to $5$ beliefs is comparable to that from $5$ to $10$ beliefs, suggesting a possible saturation value, although the precise saturation point remains undetermined. Ring agents exhibit the anticipated convergence of all results toward the triangle case, except for the dynamical transition, which occurs at a temperature about $0.1$ units lower for the triangle. For star-like agents, the dynamical transition occurs at the same temperature for the central belief $A$ and the peripheral belief $B$, and the curves overlap for the 3-node open chain and the 4-node star. As $c$ increases, the jump from neutral to extremist consensus happens at higher temperatures; similar to the clique agents, the results suggest that this dynamical-transition temperature may saturate at a certain value of $c$.

\section{Zero temperature behavior}\label{sec:zero}
\begin{figure}
\centering
\includegraphics[width=0.95\columnwidth]{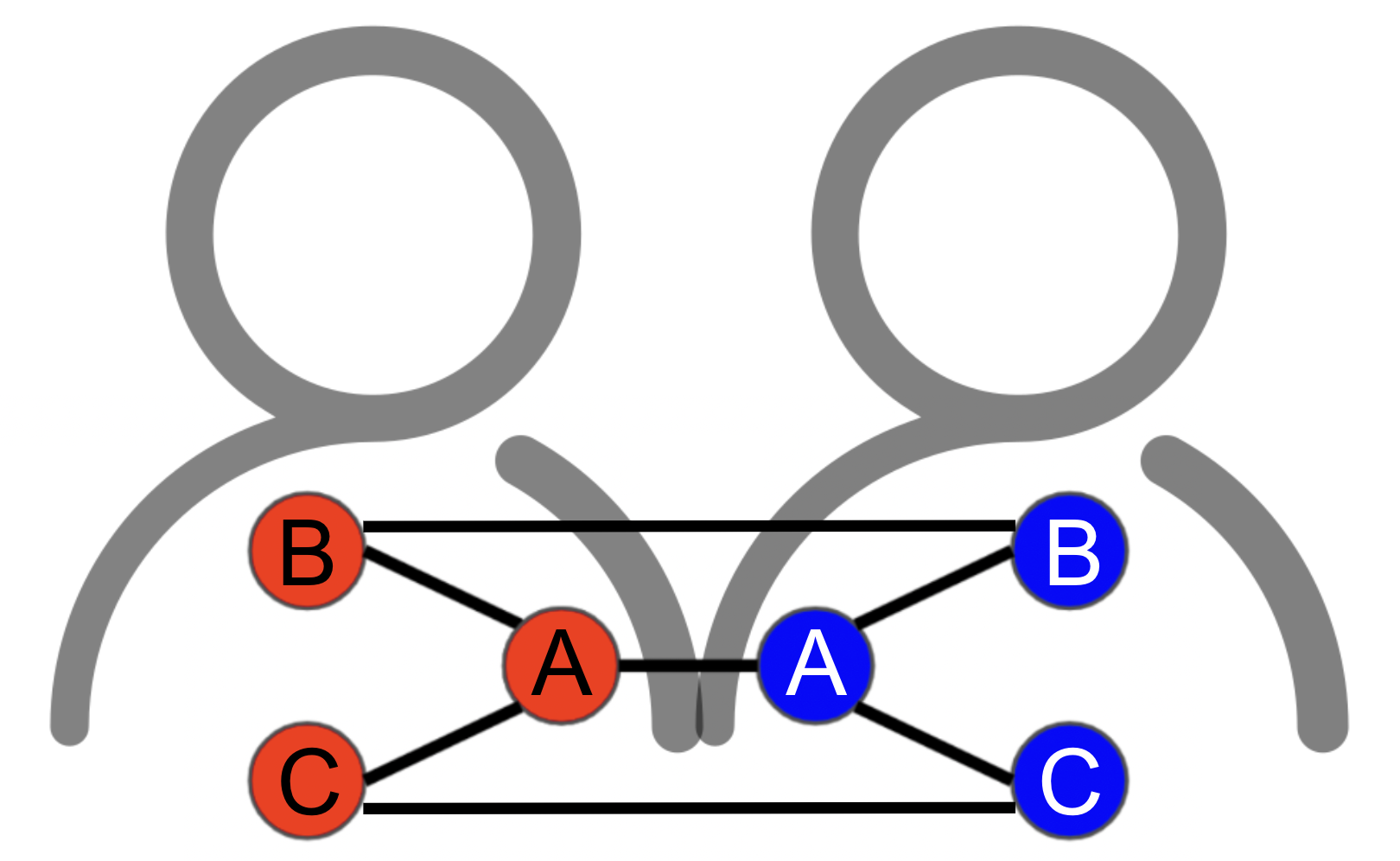}
\caption{\small{Example of two star-like agents with three in-
ternal beliefs, each in internal coherence but at opposite
states. Red denotes the $-1$ state and blue denotes the $+1$
state.}}
\label{Fig:T0_example}
\end{figure}

At zero temperature, the system becomes trapped in a collection of metastable states whose nature depends on the system’s parameters. For $\alpha = 0.85$, all beliefs reach the central state, leading to a global neutral consensus. However, at $\alpha = 0$, the system becomes trapped in states with mixtures of extremist nodes, akin to the behavior seen in Erd\H{o}s--R\'enyi graphs \cite{Ferri_2022}. Clique and star agents reach steady states with two factions of agents who are internally coherent but hold opposite extremist opinions, characterized by internal energy $E_{\mathrm{int}} = -c/2$ (see Fig.~\ref{Fig:T0_example}). Once internally aligned, they cannot overcome the energy barriers that would lead the dynamics to global polarized consensus. Ring agents differ because their internal topology can support domain-like configurations and internally dissonant metastable states.

\begin{figure}
\centering
\includegraphics[width=1\columnwidth]{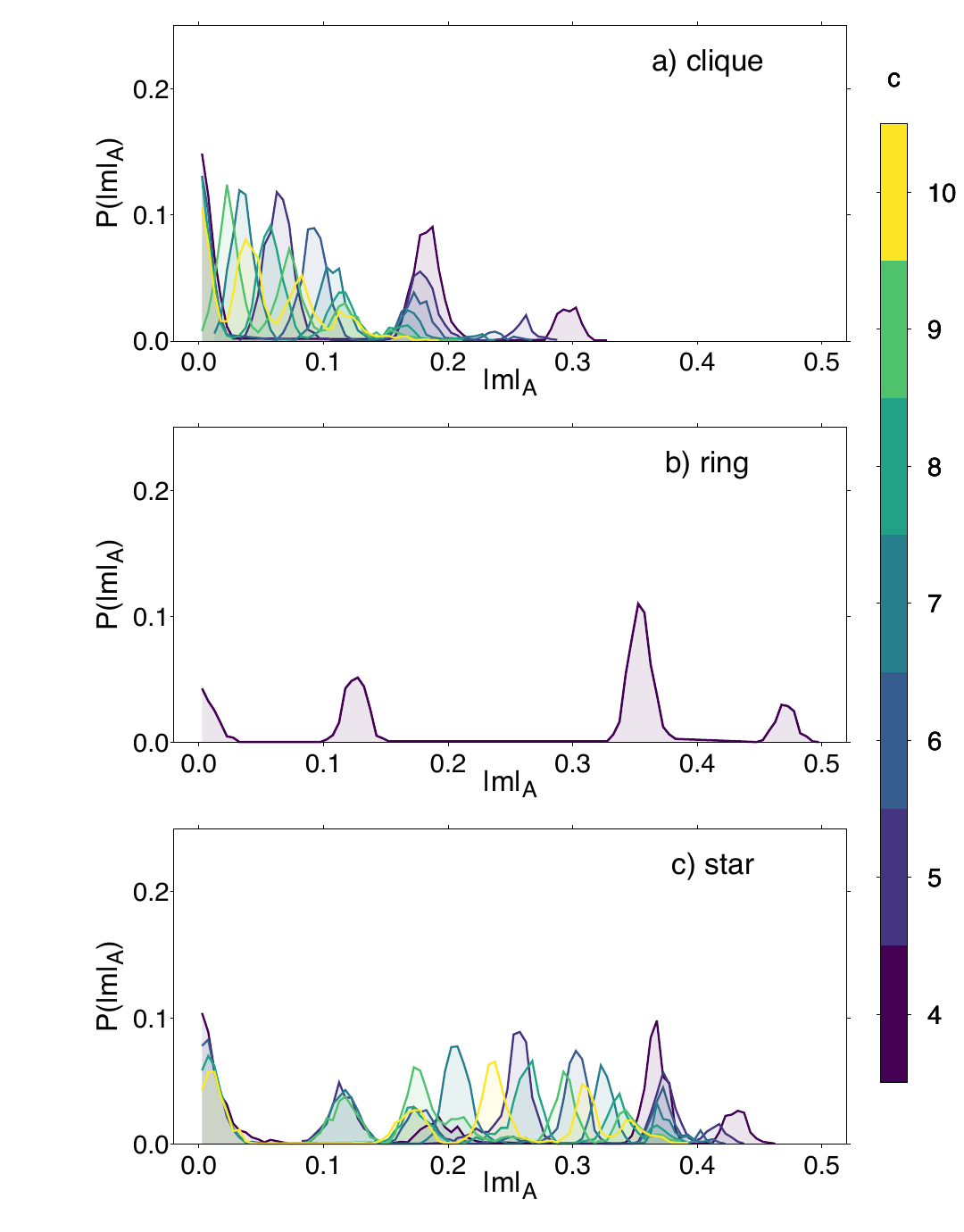}
.

\caption{\small{Zero-temperature distributions of the absolute belief-$A$ magnetization, $|m_A|$, for isolated systems of clique, ring, and star agents; colors denote the number of internal beliefs, $c = 4,\ldots,10$. Under this dynamics, all clique and star agents reach internally coherent final states for every tested $c$; in the reference case of ring agents with $c = 4$ (panel (b)), all agents reach internal agreement in about 75\% of the simulations. Larger ring agents often retain internally dissonant configurations and are therefore analyzed separately in Fig.~\ref{Fig:T0_ring}. Results are from 2000 independent runs with $N = 10^4$ agents and $\alpha = 0$.}}
\label{Fig:T0}
\end{figure}

In Fig.~\ref{Fig:T0} we present the distributions of the absolute magnetization of belief $A$, $|m_A|$, obtained from 2000 simulations for the cases shown, allowing the system to evolve for $c\cdot 10^{4}$ MC steps. Clique agents display a first peak around $|m_A| = 0$, followed by subsequent peaks of decreasing height. As $c$ increases, these later maxima shift toward lower magnetization values, suggesting that, with a sufficiently large number of beliefs, only the peak at $|m_A| = 0$ will persist, which corresponds to two opposite extremist factions of roughly equal size.

For clique and star agents, the position of the lowest-magnetization peak depends on the parity of $c$. The near-zero peak appears for the even values of $c$ examined, whereas for odd $c$ it is shifted to a small but nonzero value. This parity dependence affects the shape of the zero-temperature magnetization distribution but does not change the macroscopic interpretation of the final state as a split between opposite extremist factions.

Star agents exhibit a series of peaks similar to those of the clique agents; however, unlike cliques, their maxima do not consistently converge toward $\lvert m\rvert = 0$ when $c$ increases, nor is the tallest peak the one with the lowest magnetization value. This implies that star agents are more likely than cliques to reach a steady state in a polarized, bipartite scenario with a dominant majority of one of the two extremist factions, rather than in a situation with two equally sized opposing groups. The peaks for star agents are found within the range $\lvert m\rvert \in [0,0.5]$ and maintain a similar profile across all studied numbers of beliefs.

Since the critical temperature for the one-dimensional chain is zero \cite{Ferri_2022}, ring agents can retain internally dissonant configurations at $\alpha=0$. Because the coupling constants are normalized, this behavior is not due to the smaller number of internal links per belief \emph{per se}, but to the ring connection pattern.

\begin{figure}
  \centering

 \includegraphics[width=1\columnwidth]{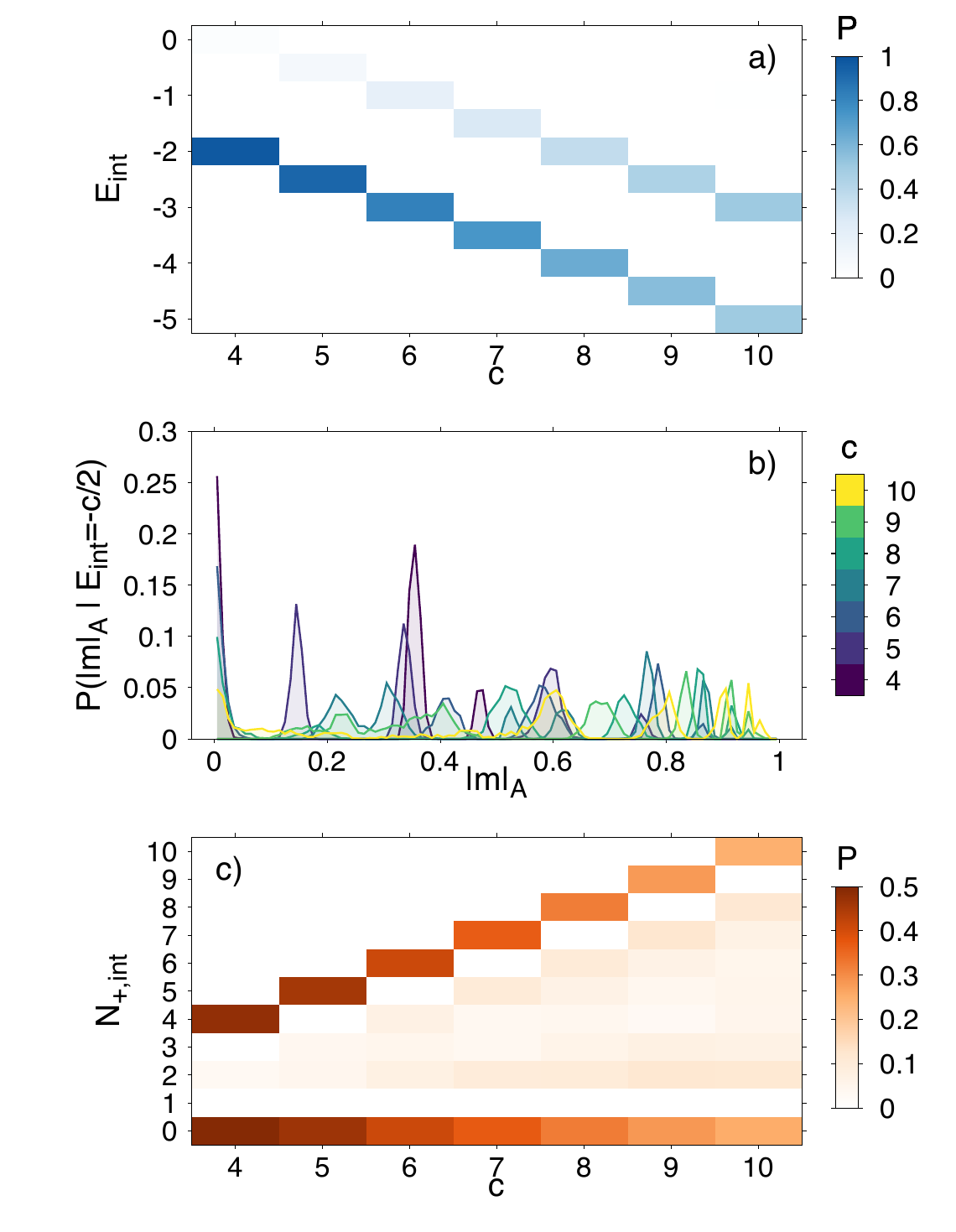}

  \caption{\small Distribution of internal parameters per agent for isolated systems of ring agents at zero temperature and $\alpha = 0$, for $c = 4,\ldots,10$. Panel (a): distribution of internal energy, $E_{\mathrm{int}}$; panel (b): distribution of the absolute magnetization of belief $A$, $|m_A|$, computed over agents that reach internal agreement, $E_{\mathrm{int}} = -c/2$; panel (c): distribution of $N_{+,\mathrm{int}}$, the number of internal beliefs in state $+1$. Results are from 2000 independent runs with $N = 10^4$ agents.}
  \label{Fig:T0_ring}
\end{figure}

For $c=4$ (see Fig.~\ref{Fig:T0}(b)), the ring distribution shows the same qualitative pattern as the $c=4$ clique and star agents: the steady state splits into two opposing extremist factions and the $\lvert m_A\rvert$ distribution is confined to low magnetization values, with no peaks at large $\lvert m_A\rvert$. As with the $c=4$ star agents, the tallest peak appears at an intermediate $\lvert m_A\rvert$ rather than at zero, so the dominant outcome involves a majority faction opposed by a smaller minority rather than two equally sized factions. We therefore present the $c=4$ ring in Fig.~\ref{Fig:T0} alongside the $c=4$ clique and star, while rings with larger $c$, where internal coherence breaks down and $\lvert m_A\rvert$ presents peaks at large values, are analyzed separately in Fig.~\ref{Fig:T0_ring}. The $c=4$ ring differs from the $c=4$ clique and star only quantitatively: about $75\%$ of $c=4$ ring realizations reach full internal agreement (compared with all realizations for cliques and stars), and in the remaining $25\%$ only about $1/8$ of the agents end in internal disagreement.

For larger $c$, a significant proportion of ring agents conclude the simulation in internally dissonant states. The internal-energy distributions in Fig.~\ref{Fig:T0_ring}(a) display a coherent peak at $E_{\mathrm{int}}=-c/2$ and a second peak at $E_{\mathrm{int}}=2-c/2$, corresponding to two internal clusters of opposing opinions, often of comparable size (see Fig.~\ref{Fig:T0_ring}(c)), separated by two domain boundaries, each contributing an energy cost of $1$. The evolution of rings toward internal agreement takes place via a domain-growth process, in which the walls between domains of different signs perform a random walk. The peak at $E_{\mathrm{int}}=-c/2$ decreases monotonically with $c$, and it is plausible that it vanishes for even larger $c$, while new peaks at higher energies may emerge due to the formation of more than two belief domains within the rings.

Fig.~\ref{Fig:T0_ring}(b) shows the distribution of the absolute magnetization of belief $A$, $|m_A|$, similar to Fig.~\ref{Fig:T0}, but computed only over ring agents that reach internal agreement, $E_{\mathrm{int}}=-c/2$. In contrast to what we observe for systems of clique and star agents, the conditioned distributions develop peaks at large $|m_A|$ as $c$ increases, corresponding to strongly polarized belief-$A$ outcomes. These distributions have noisier magnetization peaks than in systems where all agents reach internal coherence, while the peak at zero shows the same parity dependence shown in Fig.~\ref{Fig:T0}.

Panel (c) of Fig.~\ref{Fig:T0_ring} shows the distribution of $N_{+,\mathrm{int}}$, the number of internal beliefs in state $+1$; the distribution for the $-1$ state is identical by symmetry. The highest peaks are at $N_{+,\mathrm{int}}=0$ and $N_{+,\mathrm{int}}=c$, corresponding to agents in internal agreement in the two opposite extremist states. The probabilities of $N_{+,\mathrm{int}}=1$ and $N_{+,\mathrm{int}}=c-1$ are zero because configurations with a single internal belief opposed to the rest of the agent are energetically unstable. Given that the number of possible configurations compatible with two equally sized belief clusters is larger than for clusters of different sizes, the lower central probability values are not due to combinatorial probabilities. Instead, this effect arises because agents that achieve internal coherence influence other agents to adopt higher internal magnetizations. The distribution of extremist beliefs inside each agent becomes flatter as $c$ increases because the time required for internal consensus in the ring grows with the number of nodes, thus, we expect that it tends to a uniform distribution for larger $c$.

It is worth remarking that all distributions in Fig.~9 remain unchanged even when the system evolves for a number of MC steps ten times larger; the states reached by ring agents with a large number of internal beliefs are stable. The reason is that once a particular belief (such as belief $B$) achieves a majority of positive state across all agents, those agents with neighboring beliefs $A$ and $C$ also in the $+1$ state will not accept any change for belief $B$, even if the rest of their internal beliefs are in the negative state. As a consequence, the configuration becomes frozen, as happens with the agents that achieve internal coherence.

\section{Mixed topologies}\label{sec:mix}
Depending on the topology connecting their internal beliefs, a natural question arises: do agents with a specific inner topology maintain consistent behavior when interacting with different types of agents? To address this, we have conducted MC simulations, mixing two types of agents in equal proportions. Instead of the order parameter for a particular belief, we focus on the internal order parameters of the agents. We present results at temperatures near the transitions, where agents exhibit more behavioral diversity, and we choose the largest number of internal beliefs considered in previous sections, $c=10$, which presents the greater differences among internal topologies.

\subsection{Second-order transition}
\begin{figure*}
\centering
\includegraphics[width=1\textwidth]{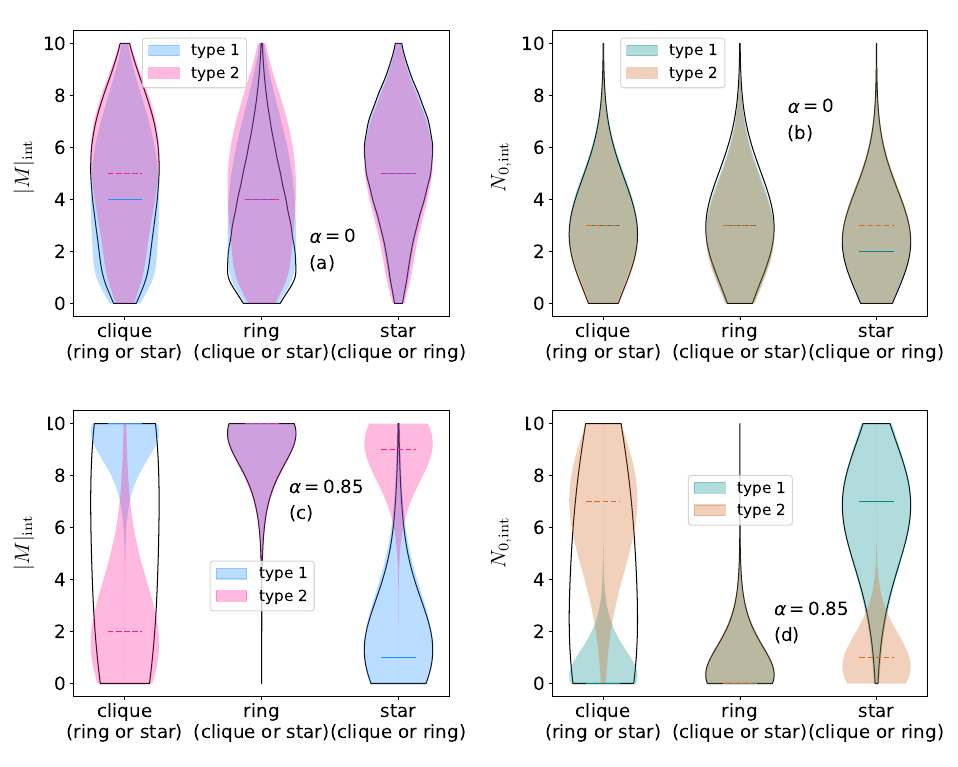}
 \caption{\small{Distribution of the absolute values of internal magnetization (first column) and the number of internal neutral beliefs (second column) for agents of each topology. The black outline corresponds to the distribution of that agent type in isolation. Colored distributions correspond to mixtures in a 50\% proportion: blue for internal magnetization (and green for neutral beliefs) when agents were mixed with the first type listed in parentheses beneath them, and in pink for internal magnetization (and orange for neutral beliefs) when mixed with the second type. Lines indicate the medians. Results were obtained from $10^4$ repetitions with systems of $N = 10^4$ agents at a temperature of $T = 1.2$ and neutrality parameter $\alpha = 0$ (top row), and at $T = 0.63$ and $\alpha = 0.85$ (bottom row).}}
\label{C6:Fig:mix}
\end{figure*}

For $\alpha=0$ (Fig.~\ref{C6:Fig:mix}, top row) we present results for $T=1.2$, which is close to the second-order transition. Distributions are generally broad, presenting all internal magnetization values, with medians at $4$ or $5$, with star agents exhibiting distributions more sharply peaked around the median. At $\alpha=0$, the distribution of neutral internal beliefs is only due to thermal fluctuations and remains approximately the same across all topologies, regardless of whether agents are isolated or mixed with another type. The median of $N_{0,\mathrm{int}}$ is $3$ ($2$ for stars mixed with rings), which is roughly $c/3$, the value expected as $T \to \infty$.

We observe that, when mixed with ring agents (blue-shaded plot), clique agents display slightly lower internal magnetizations compared to when they are alone (black line), indicating that ring agents lead to equalize the number of internal beliefs in states $+1$ and $-1$ inside the cliques. Taking into account that $N_{0,\mathrm{int}}$ is around $c/3$, this distribution of internal beliefs for clique agents is closer to having one third of their beliefs in each possible state of the model, indicative of maximal cognitive dissonance. By contrast, star agents (pink-shaded plot) lead to higher internal magnetization, promoting greater internal coherence and polarization towards one of the extreme opinions in cliques.

Rings, with their tendency to form internal clusters of opposite magnetizations of similar size, exhibit a distribution skewed towards low internal magnetization when isolated. Presumably, the neutral beliefs are situated at the boundaries between the extremist clusters, facilitating their separation, since the energy cost of a chain segment $\{+1\cdots +1,0,-1\cdots -1\}$ is equivalent to that of $\{+1\cdots +1,-1\cdots -1\}$. Both clique and star agents shift the rings towards higher $M_{\mathrm{int}}$ (although star agents have a stronger effect), pushing the rings closer to a state of maximal cognitive dissonance.

Star agents are the only ones that do not experience the influence of being mixed with another agent type. Their ``zealot'' behavior at $\alpha = 0$ remains evident in mixed simulations as the shape of their internal order parameter distribution stays consistent whether they are alone or mixed with cliques or rings. They maintain $M_{\mathrm{int}} = 5$, and the distribution is slightly skewed towards higher magnetizations, indicating a stronger internal preference for one of the extremist states.

\subsection{First-order and dynamical transitions}
For $\alpha = 0.85$ (Fig.~\ref{C6:Fig:mix}, bottom row), we chose a temperature $T = 0.63$ corresponding to the change in the basins of attraction. This represents a dynamical, not an equilibrium, effect. Here, we observe distributions that are more skewed towards one of the two limits: $0$ or $10$. The values for $N_{0,\mathrm{int}}$ are no longer merely thermal fluctuations, but complement $M_{\mathrm{int}}$, meaning that when $M_{\mathrm{int}} = 0$, it is not due to an equal number of opposite beliefs, but because there is neutral internal agreement. In this scenario, rings, rather than stars, are the dominant structure.

When cliques are alone, they display an almost flat distribution for both order parameters, indicating that any number of beliefs from $0$ to $10$ are equally likely to be in any of the three possible states. However, when mixed with rings, they become entirely extremist, with $M_{\mathrm{int}} \approx c=10$ and $N_{0,\mathrm{int}}\approx 0$. In fact, they adopt the same distribution as the rings alone. The opposite effect is observed when cliques are mixed with stars; here, cliques assume the same order-parameter distributions as the stars alone, characterized by a median $N_{0,\mathrm{int}}$ of $7$ and a low value for $M_{\mathrm{int}}$.

As mentioned before, rings, characterized by high internal magnetization with a median equal to $c=10$ and a median of zero for internal neutral beliefs, are not affected by mixing with cliques or stars. Their system maintains a ferromagnetic global phase, corresponding to the ground state of the system, despite the high value of the neutrality parameter and any perturbations caused by mixing with other agents. This stability makes them a catalyst that drives other types to converge towards the ground extremist global minima.

Star agents, when alone, have low internal magnetization values (median $=1$) and are mostly neutral. Mixing them with clique agents does not alter their internal order-parameter distributions; however, as previously mentioned, rings pull them towards extremism, completely changing their behavior.

\section{Conclusions}\label{sec:conclusions}
The systems analyzed in this study can be interpreted as a special case of multiplex networks in which each layer represents an agent and the nodes within each layer represent that agent's beliefs. Inside each layer, beliefs are connected according to the agent's internal topology, while layers are connected through same-topic interlayer links. The interpretability of this two-level structure is valuable from a qualitative perspective and can potentially be extended to other multilevel systems. Despite being nontrivial, the use of simple symmetrical structures for the internal and external networks allows for an analytical treatment of the system.

We want to remark that the topology-dependent effects reported below arise purely from the internal arrangement of beliefs, not from an imbalance between the social and internal coupling terms: the normalization factors $z_{\mathrm{ext}}$
 and $z_{\mathrm{int}}$
in the Hamiltonian ensure that both contributions scale as $\mathcal{O}(Nc)$, making them comparable.

When the neutrality parameter $\alpha$ is low, in the region of the second-order transition, adding more belief nodes to each agent generally increases the critical temperature. This increase in $T_c$ may reach saturation at an upper bound number of beliefs, depending on the topology of the internal belief network. Furthermore, additional beliefs complicate the energy landscape, steering the system dynamics toward metastable states at low temperatures, the nature of which depends on the internal topology, ranging from bipartisan deadlock with full internal coherence to partial social consensus coexisting with internal cognitive dissonance.

With high $\alpha$, on the other hand, neutral agents are not merely undecided individuals but play a significant role in the opinion-spreading process. When the neutrality parameter exceeds the tricritical point for one internal belief, all topologies undergo a first-order transition at a temperature that depends solely on $\alpha$, except for 2-belief and triangle agents whose critical temperature is slightly lower. Moreover, all systems exhibit neutral consensus as a dynamic attractor at low temperatures, but they transition to the global energy minimum corresponding to the extremist consensus at a given temperature, prior to the first-order transition. This sequence suggests that moderate levels of social agitation can drive a centrist population into collective radicalization, while excessive agitation dissolves any coherent collective stance. The pattern is consistent with Hoffer's observation that mass movements recruit most effectively from populations experiencing upheaval \cite{Hoffer_1951}, and with models showing that perturbations to initially stable belief systems can trigger abrupt societal transitions \cite{Rodriguez_2016, Kovacs_2025}. The temperature of this dynamical transition is influenced by both the internal topology and the number of beliefs.

Mixing agents with different internal topologies in the same simulation significantly impacts their behavior. The topology whose single-type system lies farther from its transition regime tends to preserve its isolated order-parameter distribution under mixing, while the other agent type is biased toward this behavior. Overall, when mixing at $\alpha = 0.85$, close to the temperature corresponding to the change in dynamical basins of attraction, the impact is more significant than when close to the second-order phase transition at $\alpha=0$. In general, we expect that the phenomena related to mixing agent types strongly depend on the selected temperature and neutrality parameter. In particular, our results suggest a crossover between regimes in which different topologies act as catalysts. Mapping the boundaries of these regimes across the $(T, \alpha)$ parameter space, as well as exploring the role of mixing proportions (for instance, whether a critical fraction of the dominant topology is required to shift the behavior of the other) remain open directions for future work.

In clique agents, all internal beliefs support each other. With increasing $c$, the order parameters approach a saturated regime where agents act as a supernode. Their beliefs are tightly coherent, functioning as a single unit, leading to an extremist bipartisan situation at zero temperature and low $\alpha$.

Rings with any number of internal beliefs are effectively equivalent to triangles at equilibrium. However, at low temperatures, for $\alpha = 0$ and large $c$ values, they exhibit a more complex behavior.
The sparse support structure of ring topologies, where each belief is reinforced by only two neighbors rather than the full set, leads to metastable configurations in which agents struggle to reach internal coherence. Among those agents that do achieve internal coherence, the dynamics can lead to bipartisan deadlock (as cliques and stars do), but uniquely for rings, also to near-complete extremist social consensus on one or more specific beliefs.
In other words, the difficulty in reaching internal consensus enhances the influence of peer pressure, as agents who have aligned internally push the system towards a global polarized consensus for some beliefs (consistent with arguments in previous sociological studies \cite{Zaller_1992, Baldassarri_2007}) while leaving the remaining beliefs unresolved for a subset of agents trapped in persistent internal cognitive dissonance.

Star agents, with a large enough number of peripheral beliefs supporting their core beliefs, act as zealots and maintain a global extremist consensus phase that is difficult to destabilize. Even when mixed with other agent types near the transition temperature, they remain uninfluenced by their presence, retaining the same internal order parameters as if isolated. However, this is only true for low $\alpha$. As neutral beliefs gain strength, star agents undergo a first-order phase transition just at the same temperature as cliques or rings. Moreover, during the dynamic shift from neutral to extremist consensus, it is the star agents that are influenced by the rings, which dominate the dynamics when mixed with other types in this parameter region.

The multilayer approach offers a versatile framework for analyzing social–belief systems from both bottom-up and top-down perspectives. In a bottom-up approach, researchers can examine the topology of belief networks within population samples to infer the potential impact of their topology on the social dynamics and collective behavior of the aforementioned population. On the other hand, the top-down approach could be used to understand psychological states such as cognitive dissonance through a social analysis of an individual’s network of social contacts.
Network representations and interpretations of belief systems are a current research line in psychology, and as a result, the availability of data for studying such networks is steadily increasing \cite{Dalege_2017_Attitudes, Borsboom_2021_NetworkAnalysis}.
It could be interesting to adapt traditional psychological questionnaires and fine-tune them for a more accurate testing of the model.

Besides its potential social interpretation, this study’s results are innovative in the realm of Ising-based models on multilayer networks. Although some previous works have explored this area \cite{Gomez_2015, Jang_2015, Chmiel_2017, Krawiecki_2018, Zhang_2023}, the existing body of literature remains limited. Expanding this research to include other structures for the external network, such as scale-free or modular graphs, could provide insights into their impact on the phase diagram and the system’s dynamical attractors. Furthermore, incorporating asymmetric, weighted, or complex adjacency matrices and studying other dynamics could yield valuable findings with broad applicability.

To summarize, the multiplex network-of-networks representation used here is still in an early stage of development and holds great potential for application across fields ranging from biological to technical systems.
Despite significant simplifications in the external and internal network structures, this work is expected to serve as a useful tool for understanding nested complex networks that resemble multiscale system structures \cite{Arenas_2008_mar, SalesPardo_2007_Hierarchical}.
Noticeably, the separation of timescales between the internal and social dynamics is not imposed by design in this model; rather, depending on the topology, temperature, and neutrality parameter, the two levels may effectively decouple, with internal consensus freezing before social consensus is reached, or co-evolve on comparable timescales. This emergent timescale separation is a feature that further extensions to other dynamics and topologies could help characterize more precisely, contributing to a broader understanding of the multiscale problem in complex systems. Coarse-graining is an important step in modeling, yet fine-tuning specific system characteristics can enhance our understanding of the nuances in complex systems.

\begin{acknowledgments}
\begingroup\raggedright

We acknowledge financial support from MINECO/FEDER, EU, through Project No.~PID2021-\allowbreak{}128005NB-\allowbreak{}C22 and Project No.~PID2024-\allowbreak{}158120NB-\allowbreak{}C22; from MCIN/AEI/\allowbreak{}10.13039/\allowbreak{}501100011033 through Grant No.~PRE2019-\allowbreak{}090279; and from the Generalitat de Catalunya through Grant No.~2021SGR00856. We also acknowledge support from the NSF AccelNet-MultiNet program, funded by the National Science Foundation under Award Nos.~1927425 and~1927418.
\par\endgroup
\end{acknowledgments}

\bibliographystyle{apsrev4-2}
\bibliography{Bibliography_Books.bib, Bibliography.bib, ifo_bib.bib, mobile.bib}

\end{document}